\documentstyle[twoside,epsfig]{article}

\catcode`\@=11
\long\def\@makefntext#1{
\protect\noindent \hbox to 3.2pt {\hskip-.9pt  
$^{{\eightrm\@thefnmark}}$\hfil}#1\hfill}		%CAN BE USED 

\def\@makefnmark{\hbox to 0pt{$^{\@thefnmark}$\hss}}	%ORIGINAL 
	
\def\ps@myheadings{\let\@mkboth\@gobbletwo
\def\@oddhead{\hbox{}
\rightmark\hfil\eightrm\thepage}   
\def\@oddfoot{}\def\@evenhead{\eightrm\thepage\hfil
\leftmark\hbox{}}\def\@evenfoot{}
\def\sectionmark##1{}\def\subsectionmark##1{}}

\oddsidemargin=\evensidemargin
\newcounter{sectionc}\newcounter{subsectionc}\newcounter{subsubsectionc}
\renewcommand{\section}[1] {\vspace{12pt}\addtocounter{sectionc}{1} 
\setcounter{subsectionc}{0}\setcounter{subsubsectionc}{0}\noindent 
	{\tenbf\thesectionc. #1}\par\vspace{5pt}}
\renewcommand{\subsection}[1] {\vspace{12pt}\addtocounter{subsectionc}{1} 
	\setcounter{subsubsectionc}{0}\noindent 
	{\bf\thesectionc.\thesubsectionc. {\kern1pt \bfit #1}}\par\vspace{5pt}}
\renewcommand{\subsubsection}[1] {\vspace{12pt}\addtocounter{subsubsectionc}{1}
	\noindent{\tenrm\thesectionc.\thesubsectionc.\thesubsubsectionc.
	{\kern1pt \tenit #1}}\par\vspace{5pt}}

\newcounter{appendixc}
\newcounter{subappendixc}[appendixc]
\newcounter{subsubappendixc}[subappendixc]
\renewcommand{\thesubappendixc}{\Alph{appendixc}.\arabic{subappendixc}}
\renewcommand{\thesubsubappendixc}
	{\Alph{appendixc}.\arabic{subappendixc}.\arabic{subsubappendixc}}

\renewcommand{\appendix}[1] {\vspace{12pt}
        \refstepcounter{appendixc}
        \setcounter{figure}{0}
        \setcounter{table}{0}
        \setcounter{lemma}{0}
        \setcounter{theorem}{0}
        \setcounter{corollary}{0}
        \setcounter{definition}{0}
        \setcounter{equation}{0}
        \renewcommand{\thefigure}{\Alph{appendixc}.\arabic{figure}}
        \renewcommand{\thetable}{\Alph{appendixc}.\arabic{table}}
        \renewcommand{\theappendixc}{\Alph{appendixc}}
        \renewcommand{\thelemma}{\Alph{appendixc}.\arabic{lemma}}
        \renewcommand{\thetheorem}{\Alph{appendixc}.\arabic{theorem}}
        \renewcommand{\thedefinition}{\Alph{appendixc}.\arabic{definition}}
        \renewcommand{\thecorollary}{\Alph{appendixc}.\arabic{corollary}}
        \renewcommand{\theequation}{\Alph{appendixc}.\arabic{equation}}
        \noindent{\tenbf Appendix \theappendixc #1}\par\vspace{5pt}}
\newcommand{\subappendix}[1] {\vspace{12pt}
        \refstepcounter{subappendixc}
        \noindent{\bf Appendix \thesubappendixc. {\kern1pt \bfit #1}}
	\par\vspace{5pt}}
\newcommand{\subsubappendix}[1] {\vspace{12pt}
        \refstepcounter{subsubappendixc}
        \noindent{\rm Appendix \thesubsubappendixc. {\kern1pt \tenit #1}}
	\par\vspace{5pt}}

\newcommand{\textlineskip}{\baselineskip=13pt}
\newcommand{\smalllineskip}{\baselineskip=10pt}

\def\eightcirc{
\begin{picture}(0,0)
\put(4.4,1.8){\circle{6.5}}
\end{picture}}
\def\eightcopyright{\eightcirc\kern2.7pt\hbox{\eightrm c}} 

\newcommand{\copyrightheading}[1]
	{\vspace*{-2.5cm}\smalllineskip{\flushleft
	{\footnotesize International Journal of Modern Physics A, #1}\\
	{\footnotesize $\eightcopyright$\, World Scientific Publishing
	 Company}\\
	 }}

\def\abstracts#1#2#3{{
	\centering{\begin{minipage}{4.5in}\baselineskip=10pt\footnotesize
	\parindent=0pt #1\par 
	\parindent=15pt #2\par
	\parindent=15pt #3
	\end{minipage}}\par}} 

\renewenvironment{thebibliography}[1]
	{\frenchspacing
	 \ninerm\baselineskip=11pt
	 \begin{list}{\arabic{enumi}.}
	{\usecounter{enumi}\setlength{\parsep}{0pt}
	 \setlength{\leftmargin 12.7pt}{\rightmargin 0pt} %FOR 1--9 ITEMS
	 \setlength{\itemsep}{0pt} \settowidth
	{\labelwidth}{#1.}\sloppy}}{\end{list}}

\newcounter{itemlistc}
\newcounter{romanlistc}
\newcounter{alphlistc}
\newcounter{arabiclistc}

\newcommand{\fcaption}[1]{
        \refstepcounter{figure}
        \setbox\@tempboxa = \hbox{\footnotesize Fig.~\thefigure. #1}
        \ifdim \wd\@tempboxa > 5in
           {\begin{center}
        \parbox{5in}{\footnotesize\smalllineskip Fig.~\thefigure. #1}
            \end{center}}
        \else
             {\begin{center}
             {\footnotesize Fig.~\thefigure. #1}
              \end{center}}
        \fi}

\newcommand{\tcaption}[1]{
        \refstepcounter{table}
        \setbox\@tempboxa = \hbox{\footnotesize Table~\thetable. #1}
        \ifdim \wd\@tempboxa > 5in
           {\begin{center}
        \parbox{5in}{\footnotesize\smalllineskip Table~\thetable. #1}
            \end{center}}
        \else
             {\begin{center}
             {\footnotesize Table~\thetable. #1}
              \end{center}}
        \fi}

\def\@citex[#1]#2{\if@filesw\immediate\write\@auxout
	{\string\citation{#2}}\fi
\def\@citea{}\@cite{\@for\@citeb:=#2\do
	{\@citea\def\@citea{,}\@ifundefined
	{b@\@citeb}{{\bf ?}\@warning
	{Citation `\@citeb' on page \thepage \space undefined}}
	{\csname b@\@citeb\endcsname}}}{#1}}

\newif\if@cghi
\def\cite{\@cghitrue\@ifnextchar [{\@tempswatrue
	\@citex}{\@tempswafalse\@citex[]}}
\def\citelow{\@cghifalse\@ifnextchar [{\@tempswatrue
	\@citex}{\@tempswafalse\@citex[]}}
\def\@cite#1#2{{$\null^{#1}$\if@tempswa\typeout
	{IJCGA warning: optional citation argument 
	ignored: `#2'} \fi}}

\def\pmb#1{\setbox0=\hbox{#1}
	\kern-.025em\copy0\kern-\wd0
	\kern.05em\copy0\kern-\wd0
	\kern-.025em\raise.0433em\box0}

\def\fnt#1#2{\footnotetext{\kern-.3em
	{$^{\mbox{\scriptsize #1}}$}{#2}}}

\def\fpage#1{\begingroup
\voffset=.3in
\thispagestyle{empty}\begin{table}[b]\centerline{\footnotesize #1}
	\end{table}\endgroup}

\def\runninghead#1#2{\pagestyle{myheadings}
\markboth{{\protect\footnotesize\it{\quad #1}}\hfill}
{\hfill{\protect\footnotesize\it{#2\quad}}}}
\font\tenrm=cmr10
\font\tenit=cmti10 
\font\tenbf=cmbx10
\font\bfit=cmbxti10 at 10pt
\font\ninerm=cmr9

\font\eightrm=cmr8

\def\qed{\hbox{${\vcenter{\vbox{			%HOLLOW SQUARE
   \hrule height 0.4pt\hbox{\vrule width 0.4pt height 6pt
   \kern5pt\vrule width 0.4pt}\hrule height 0.4pt}}}$}}

\begin{document}
\addtolength{\textheight}{.9cm}
\runninghead{Top Production and Decay at Linear Colliders
 $\ldots$} {Top Production and Decay at Linear Colliders $\ldots$}

\normalsize\textlineskip
\thispagestyle{empty}
\setcounter{page}{1}

\copyrightheading{}			%{Vol. 0, No. 0 (1993) 000--000}

\vspace*{0.88truein}
\vspace*{-9\baselineskip}
\begin{flushright}
\hfill {\rm UR-1619}\\
\hfill {\rm ER/40685/957}\\
\hfill {\rm December 2000}\\
\end{flushright}
\vspace*{2\baselineskip}
\fpage{1}
\centerline{\bf TOP PRODUCTION AND DECAY AT LINEAR COLLIDERS:}
\vspace*{0.035truein}
\centerline{\bf QCD CORRECTIONS\footnote{Presented at the 2000 
Meeting of the Division of Particles and Fields of the APS, Columbus,
OH, August 9--12, 2000.}}
\vspace*{0.3truein}
\centerline{\footnotesize COSMIN MACESANU and LYNNE H. ORR}
\vspace*{0.015truein}
\centerline{\footnotesize\it Department of Physics and Astronomy, 
University of Rochester}
\baselineskip=10pt
\centerline{\footnotesize\it Rochester NY 14627-0171, USA}
%\vspace*{0.225truein}
%\publisher{(received date)}{(revised date)}

\vspace*{0.21truein}
\abstracts{We present the results of an exact calculation of gluon 
radiation in top production and decay at high energy
electron-positron colliders. We include all spin 
correlations and interferences, the bottom quark mass, and
finite top width effects in the matrix element calculation. 
We study properties of the radiated gluons and
implications for top mass measurement. 
We also discuss virtual corrections to the process. }{}{}

%\textlineskip			%) USE THIS MEASUREMENT WHEN THERE IS
%\vspace*{12pt}			%) NO SECTION HEADING

\vspace*{1pt}\textlineskip	%) USE THIS MEASUREMENT WHEN THERE IS
\section{Introduction}	%) A SECTION HEADING
\vspace*{-0.5pt}
\noindent
Lepton colliders provide a clean environment for top quark studies.
In constrast to hadron colliders, kinematics and QCD backgrounds
are well under control at lepton colliders.  This allows for measurements
of top quark properties such as mass, width, and electromagnetic couplings.
These measurements require precise predictions, beyond leading order
in perturbation theory.  Furthermore, because the top quark
is a heavy, unstable particle, higher-order corrections to both 
the top production and decay processes must be considered together.
This is becoming the standard approach as we enter the TeV regime and 
is already familiar from dealing with $Z$ and $WW$  production at LEP.

Here we consider QCD  corrections to $t\bar{t}$ production 
and decay at linear colliders at energies above the top pair
threshold\cite{mo}; see also \cite{schmidt}. 
For most of the talk we consider real gluon corrections, and comment 
briefly on virtual corrections at the end.

\section{Real Gluon Radiation} 
\noindent
Radiated gluons appear as hadronic
jets in particle detectors, and these jets are often indistinguishable from
the jets coming from top quark decays (indeed, the gluons may be part
of the decays themselves).  As a result, identification of top quark events
and  measurement  of the top mass by momentum reconstruction can be 
complicated by the presence of gluon radiation.

At lepton colliders there can be no gluon radiation from the initial
state, but gluons can be radiated from the produced quarks in either
the top production or decay stage.  We have calculated the cross section for 
the process
\begin{equation}
e^+e^- \rightarrow \gamma^*, Z^* \rightarrow t\bar{t} (g)
\rightarrow bW^+ \bar{b}W^-g\; .
\end{equation}
using the exact matrix elements including top width effects, spin 
correlations between top production and decay,  all masses, and
all interferences\cite{mo}.

The distinction between production- and decay-stage emissions
is based on whether the top quark is closer to being on shell
before or after the top quark is radiated.  In the calculation
the separation is straightforward with a little algebra; we can
then identify the individual contributions from the production and 
decay stages as well as their interferences (which are small for
the most part).
\begin{figure*}[t]
\vskip -.5cm
\psfig{figure=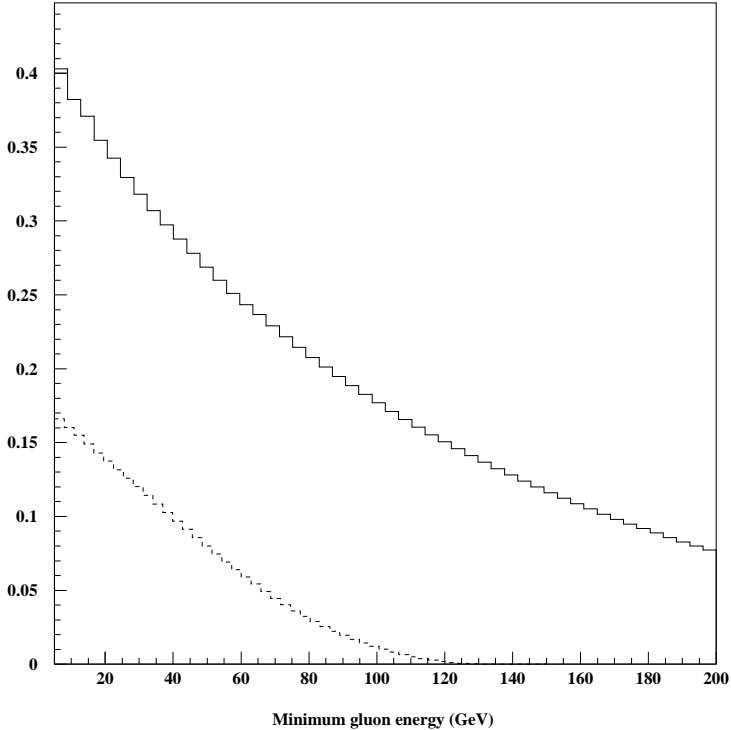,width=11cm}
\vskip -.5cm
\caption[]{  
The fraction of gluon emissions radiated in the production stage, as
a function of minimum gluon energy, for center-of-mass energy 1 TeV (solid
line) and 500 GeV (dashed line).\label{prodfrac}}
\end{figure*}

We show in Figure 1 the resulting fraction of the cross section contributed by
production-stage radiation 
as a function of the minimum gluon energy.  We see that gluons emitted in
the production stage account for less than half of the radiative cross 
section in both cases.  Phase space accounts both for the increase
in the fraction with center of mass energy and the decrease with 
increasing gluon energy.

Results for mass reconstruction are shown in Figure 2, with and without
the extra gluon included.  We see clear peaks where the correct
combination is obtained, with a low tail when decay-stage gluons 
are omitted, and a high tail where production-stage gluons are included.
\begin{figure*}[t]
\vskip -.5cm
\psfig{figure=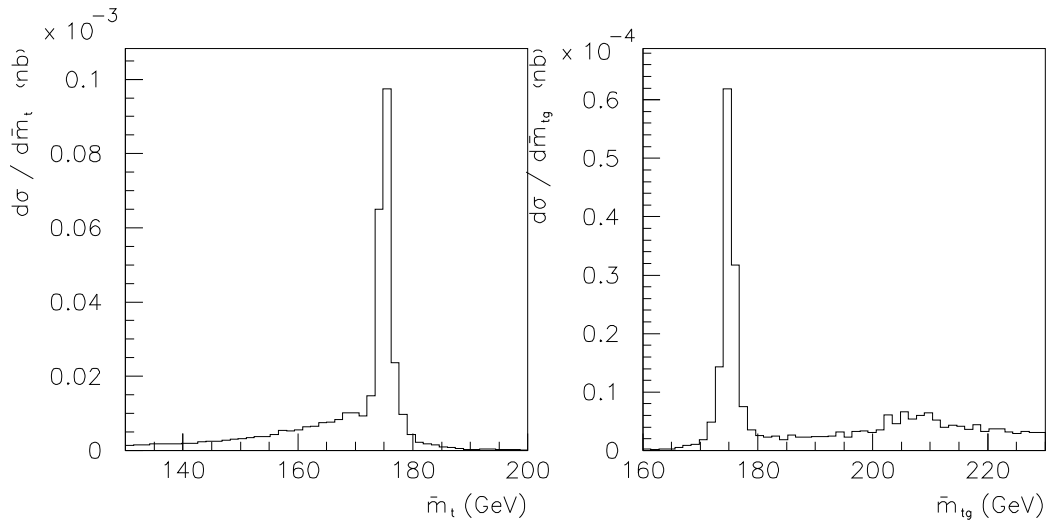,width=11cm}
\vskip -.75cm
\caption[]{  The  top invariant mass spectrum without (left) and with (right) 
the gluon momentum included, for center-of-mass energy 600 GeV.
\label{masses}}
\end{figure*}
An invariant mass cut can be used to improve the resolution and angular
cuts (between the gluon and the $b$ quarks) can do even better.  Our
results are at the parton level however, so hadronization and 
detector effects must ultimately be taken into account.

The interference between gluons radiated in the production and 
decay stages, though small, can be interesting
because it is sensitive to the total width of the top quark.  In
particular, this interference is largest when
the gluon energy is comparable to the top width, and given suitable
cuts, can be destructive.  See \cite{mo} for details.

\section{Virtual Corrections}
\noindent
A complete next-to-leading order calculation requires virtual corrections,
and these are currently in progress\cite{mac}.  Individual corrections
to production\cite{prod} and decay\cite{decay} (so-called ``factorizable''
corrections) have been known for some time.  But the interference-type
or ``nonfactorizable'' corrections, where for example a gluon connects the 
top quark and the bottom antiquark, have not.  Many of the theoretical
issues are similar to those that arise in the calculation of corrections
to $WW$ production at LEP\cite{WW}.  As in the $WW$ case, the 
calculation is being performed in the Double-Pole Approximation, in which
only contributions from doubly-resonant diagrams are kept.  The complete
calculation will be aviailable soon.

\section{Acknowledgments}
\noindent
Work supported in part by the U.S. Department of Energy and the U.S. 
National Science Foundation,
under grants DE-FG02-91ER40685 and  PHY-9600155.

\end{document}